\documentclass[runningheads]{llncs}
\usepackage[T1]{fontenc}
\usepackage{graphicx}

\usepackage{enumitem}

\usepackage{subcaption}  
\usepackage{float} 

\usepackage{amsmath}
\usepackage{amssymb}  

\usepackage[table]{xcolor}

\usepackage{bbding}

\usepackage{array}

\usepackage{multirow}

\usepackage{ulem}

\usepackage{amsmath}

\usepackage{booktabs}

\usepackage{float}

\usepackage{hyperref}
\usepackage{color}

\usepackage[
    letterpaper,
    top=2cm,
    bottom=2cm,
    left=3cm,
    right=3cm,
    marginparwidth=1.75cm
]{geometry}

\begin{document}
\title{
    QvTAD: Differential Relative Attribute Learning 
    \\
    for Voice Timbre Attribute Detection
}

%
%
\author{
    Zhiyu Wu
    \and
    Jingyi Fang
    \and
    Yufei Tang
    \and
    \\
    Yuanzhong Zheng
    \and
    Yaoxuan Wang
    \and
    Haojun Fei
}

\authorrunning{Zhiyu Wu et al.}

\institute{
    Qifu Technology, Shanghai, China  \\
    \email{\{wuzhiyu, fangjingyi, tangyufei, zhengyuanzhong,  wangyaoxuan, feihaojun\}-jk@360shuke.com} 
}
\maketitle              
\begin{abstract}
Voice Timbre Attribute Detection (vTAD) plays a pivotal role in fine-grained timbre modeling for speech generation tasks. However, it remains challenging due to the inherently subjective nature of timbre descriptors and the severe label imbalance in existing datasets. In this work, we present QvTAD, a novel pairwise comparison framework based on differential attention, designed to enhance the modeling of perceptual timbre attributes.
To address the label imbalance in the VCTK-RVA dataset, we introduce a graph-based data augmentation strategy that constructs a Directed Acyclic Graph and employs Disjoint-Set Union techniques to automatically mine unobserved utterance pairs with valid attribute comparisons. Our framework leverages speaker embeddings from a pretrained FACodec, and incorporates a Relative Timbre Shift-Aware Differential Attention module. This module explicitly models attribute-specific contrasts between paired utterances via differential denoising and contrast amplification mechanisms.
Experimental results on the VCTK-RVA benchmark demonstrate that QvTAD achieves substantial improvements across multiple timbre descriptors, with particularly notable gains in cross-speaker generalization scenarios.


\end{abstract}
\section{Introduction}
Timbre attribute modeling is critical for advancing speech generative technologies. As a core element of conversation, a speaker’s timbre conveys both identity-related characteristics and unique vocal style—shaped by vocal tract anatomy, age, emotion, and environment. These variations shape auditory perception and generate diverse expressive impressions. For generative tasks like speech synthesis, developing highly expressive, finely controllable multi-speaker systems requires multi-dimensional objective modeling of synthesized timbre attributes. However, modeling timbre variations across styles remains challenging due to the subjective and abstract nature of timbre descriptions.

Some speech generation algorithms, such as Text-to-Speech (TTS) and Voice Conversion (VC), have seen studies conduct in-depth exploration into the description and extraction of speakers' timbre features. Seed-VC \cite{seedvc} enhances timbre representation using a diffusion-based Transformer architecture, while introducing training-time perturbations to mitigate timbre leakage, thereby achieving more accurate voice conversion. In the TTS domain, NaturalSpeech3 \cite{naturalspeech3} proposed a codec-based speech representation that decomposes waveform signals into prosody, linguistic content, acoustic details, and timbre of speaker. This disentangled representation improves the naturalness and fidelity of reference speaker synthesis. Meanwhile, controllable TTS via instruction-based conditioning has gained increasing attention, enabling control over emotion, speed, and energy through natural language prompts, as demonstrated by \cite{instructspeech}\cite{instructts}\cite{emovoice}. Despite these advances, fine-grained modeling of timbre remains highly challenging, and tasks such as precise timbre analysis and controllable timbre generation are still in their early stages of exploration \cite{cosyvoice3}\cite{unispeaker}.

To enable timbre-aware editing and comparison, several recent studies have attempted to capture and manipulate differences in speaker timbre attributes. For example, VoxEditor \cite{vTAD-dataset} introduces a manually annotated dataset, VCTK-RVA, which encodes relative attribute differences between speakers to support text-guided voice attribute editing. This dataset provides a controllable and reproducible benchmark for timbre modeling. Building on this foundation, the newly proposed task of vTAD, as detailed in \cite{challenge} and \cite{introduce}, focuses on pairwise comparisons of timbre attribute strengths based on the VCTK-RVA dataset. Details regarding the dataset and task setup are introduced in Section 4.1.1 and Section 3.1.1, respectively.

In this paper, we propose a novel differential attention-based pairwise comparison framework for more precise speaker timbre modeling. First, based on extensive analysis of the VCTK-RVA dataset, we design an automated graph-based data augmentation method to address imbalanced timbre attribute label distribution. Second, we leverage a pretrained speech codec model for speaker embedding extraction and introduce the RTSA$^2$(Relative Timbre Shift-Aware Differential Attention) module—a differential denoising mechanism—to learn attribute-specific contrasts between utterances from different speakers. Finally, we train the model on VCTK-RVA to predict which utterance exhibits a stronger presence of a given timbre attribute, with comparative experiments and ablation studies demonstrating improved accuracy and generalization to out-of-set scenarios.

Our main contributions are:

\begin{itemize}
\item We propose an automated graph-based data augmentation method, which constructs a Directed Acyclic Graph (DAG) via Disjoint-Set Union over VCTK-RVA to mitigate timbre attribute label imbalance.
\item We propose RTSA$^2$ structure, which models timbre embedding differences and incorporates a contrastive predictor to improve attribute strength classification accuracy both within and outside the dataset.
\item Experimental validation that our QvTAD method enhances prediction accuracy across multiple timbre descriptors, with notable gains in unseen-speaker settings.
\end{itemize}

\begin{figure}[t]
  \centering
  \begin{minipage}[b]{0.6\textwidth}
    \centering
    \includegraphics[width=\textwidth]{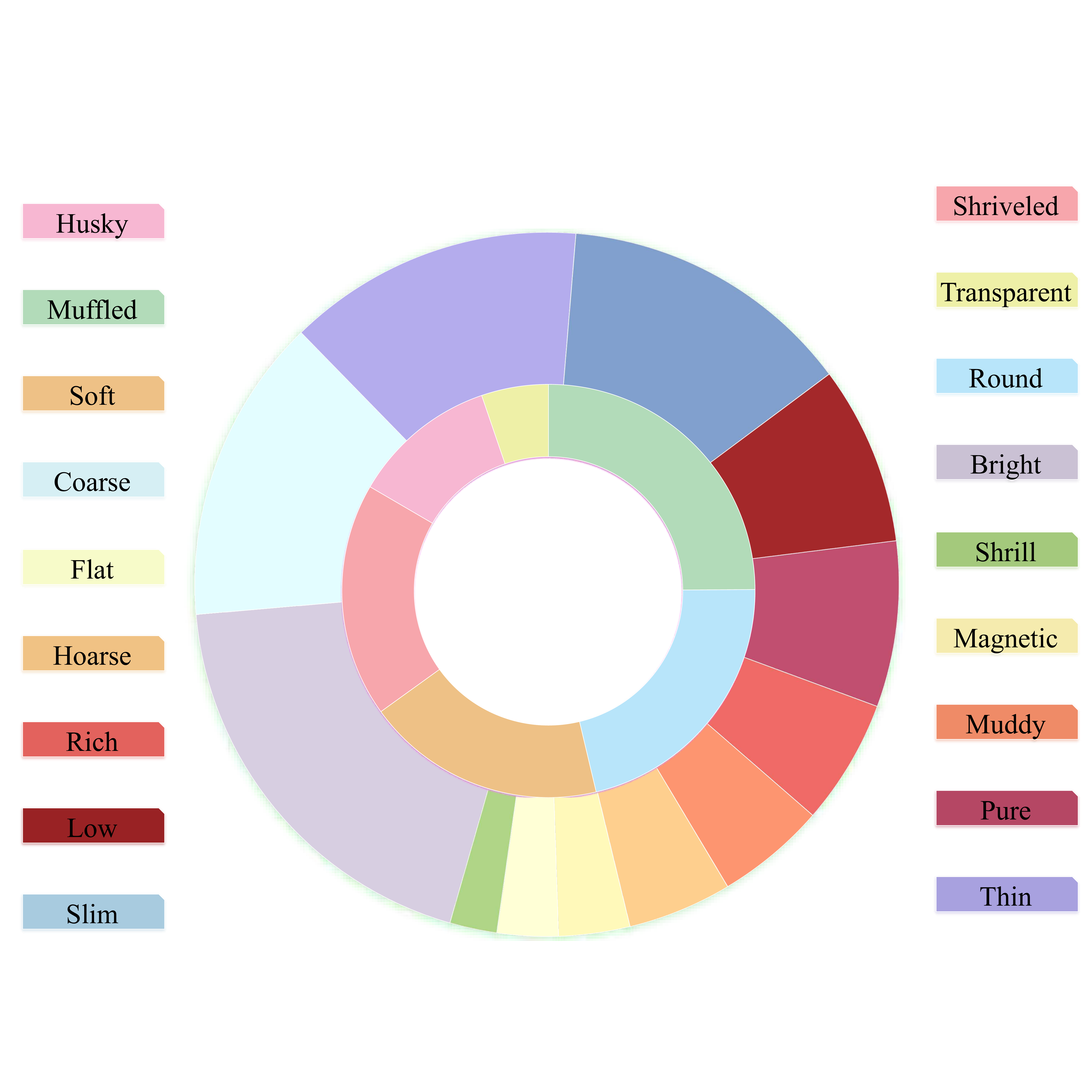}
    \subcaption{Annotated Timbre Attributes in the VCTK-RVA Dataset.} 
     \label{fig:timbre-attr} 
  \end{minipage}
  \hfill
  \begin{minipage}[b]{0.35\textwidth}
    \centering
    \includegraphics[width=\textwidth]{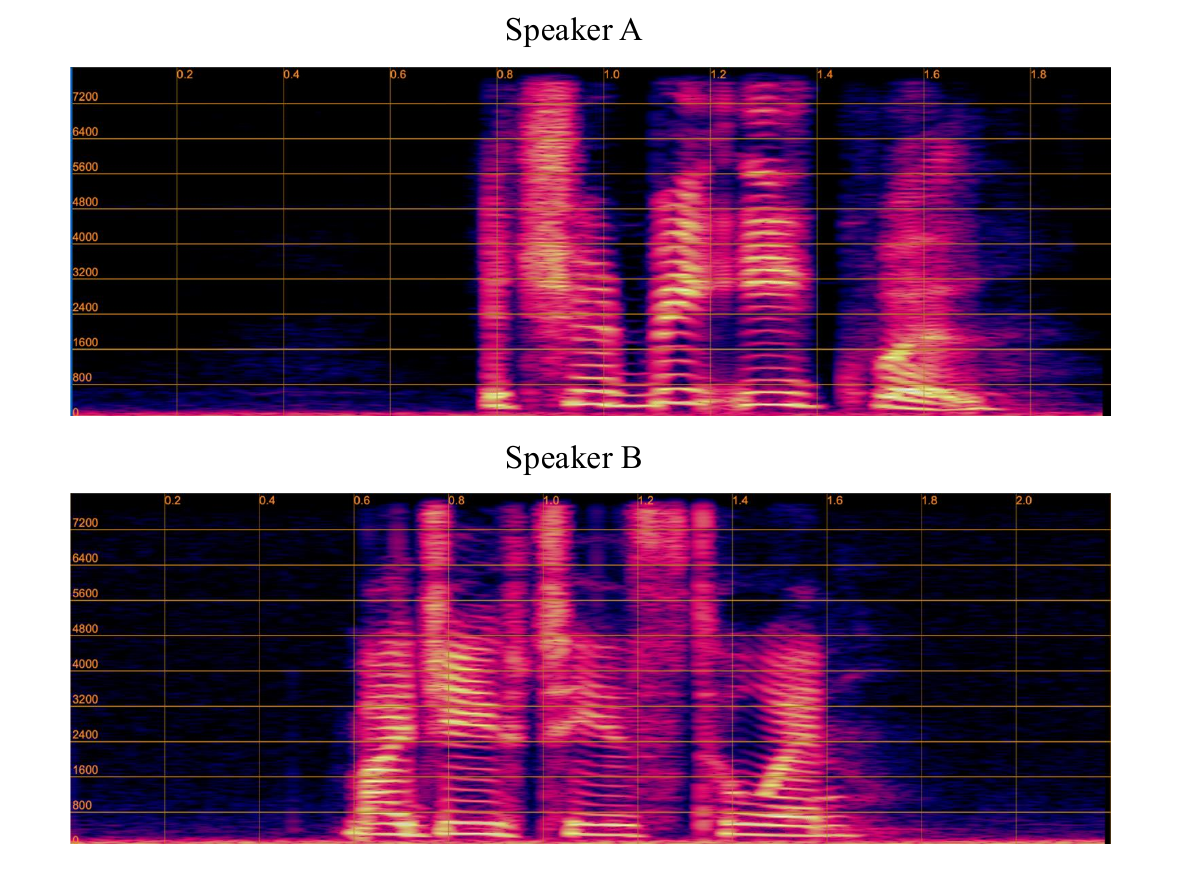}
    \subcaption{Mel-spectrogram of two speech clips from two different speakers.} 
    \label{fig:mel}
    \vspace{0.5em}
    \includegraphics[width=\textwidth]{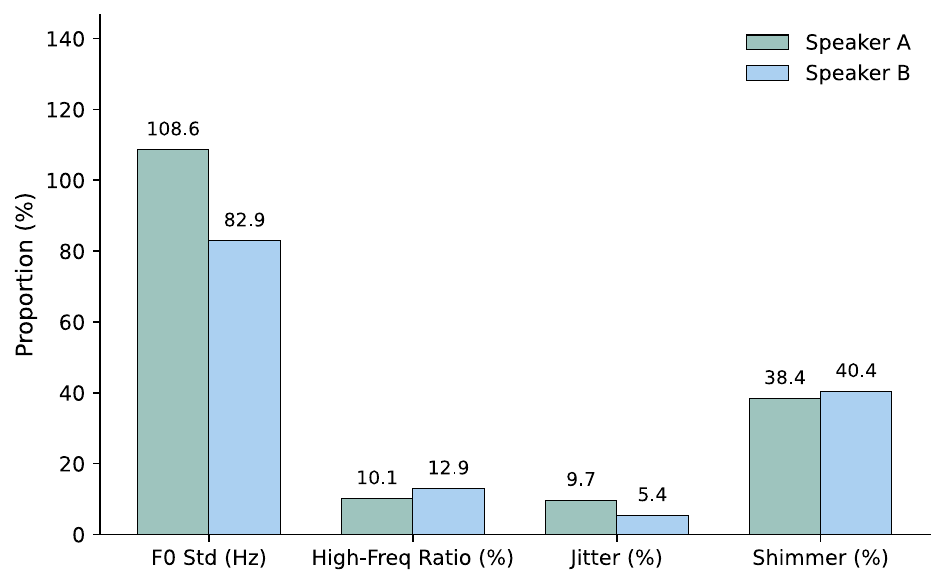}
    \subcaption{Analysis of different F0 and ratio of High frequency, Jitter and Shimmer.} 
    \label{fig:comp}
  \end{minipage}
  \caption{Introduction to Voice Attribute and case analysis of \textit{Mellow} attribute}
  \label{fig:overview}
\end{figure}

\section{Related Work}
\subsection{Cross-Domain Relative Attribute Modeling}

The task of "given two samples and an attribute, determine which sample is stronger in that attribute" is prevalent across various domains and is widely known as relative attribute learning. This formulation avoids the subjectivity and inconsistency of scalar annotations by instead supervising models through comparisons between pairs of samples. It has been applied successfully in Computer Vision, Natural Language Processing, and medical imaging, etc.

In Computer Vision, Parikh~\cite{relative_attribute} first introduced the concept of relative attributes using image pairs to define attribute strength relationships, and trained a ranking SVM accordingly. Souri proposed Deep Relative Attributes \cite{deep_relative_attribute}, leveraging CNNs for feature extraction and optimizing a ranking loss in an end-to-end framework. Zhang \cite{icml_relative_attr} introduced DACRL, incorporating attribute centers in the embedding space to improve ranking interpretability. Relative modeling of style or emotion intensity has gained traction in Natural Language Processing. Jhamtani \cite{shakespear} trained a style transfer model using pairs of Shakespearean and modern sentences to model stylistic strength. 

These methods share a unified formulation: sample pair + attribute + strength comparison, emphasizing perceptual-space modeling through relational supervision. Our task of predicting timbre strength in speech extends this paradigm to the auditory domain.

\subsection{Relative Perception and Timbre Attribute Detection}

While relative attribute learning is mature in Computer Vision and NLP, it remains underexplored in speech. Wester's MOSPC framework trains ranking models from subjective speech naturalness comparisons, outperforming traditional MOS in consistency and reliability~\cite{mospc}. NR-SQA introduced \textit{pairwise and triplet ranking losses} for no-reference speech quality assessment, directly modeling ordinal perceptual differences to improve generalization to synthetic and real speech~\cite{interspeech-pairwise}.

Despite these advances, speech attribute modeling faces two central challenges: (1) Abstraction: Timbre perception involves nonlinear interactions of frequency structure, harmonic content, and energy distribution, making it difficult to judge from Mel spectrograms. (2) Subjectivity: Timbre annotation requires expert knowledge and lacks consistency, limiting scalability.

Recent studies by VoxEditor~\cite{vTAD-dataset} introduced VCTK- RVA, the first timbre relative modeling dataset, featuring pairwise comparisons of 34 attributes (e.g., magnetic, bright, hoarse) annotated by human raters. Its attribute composition (Figure~\ref{fig:overview}a) makes it one of the few public resources for speech-based relative attribute learning.

However, due to the inherent abstraction and subjectivity, different listeners may interpret them differently. In practice, conventional signal-based acoustic analyses often suffer from a high false positive rate when modeling timbre attributes. For example, soft voices typically lack sharp high-frequency components and exhibit a concentrated energy distribution in the mid-to-low frequency bands, which results in a fuller, warmer perceived quality. These voices also tend to exhibit smoother spectral transitions, moderate and continuous formant spacing, and lower jitter and shimmer values.

As illustrated in Figure~\ref{fig:overview}b, the Mel-spectrograms of two speech segments from different speakers demonstrate varying degrees of the soft timbre. Speaker A shows smoother, continuous band textures in the 0–2 kHz range, whereas Speaker B displays more abrupt transitions despite similar energy levels. However, acoustic feature analysis in Figure~\ref{fig:overview}c reveals that Speaker A has a higher F0 Std and jitter, while Speaker B exhibits a higher high-frequency energy ratio and shimmer. Interestingly, a subjective Mean Opinion Score (MOS) evaluation conducted by four domain experts in speech AI consistently rated Speaker B as having a softer voice. This example highlights the inconsistency between signal-based acoustic features and human-perceived timbre, indicating potential limitations of low-level descriptors in modeling perceptual attributes.

While prior studies have explored pathological voice quality assessment by modeling a wide range of acoustic parameters~\cite{jitter_shimmer_interspeech}~\cite{lightly_jitter}, these methods are often constrained by modeling complexity and computational inefficiency. Moreover, handcrafted analysis remains sensitive to dataset bias and lacks robustness in large-scale applications.

\begin{figure}[t]
  \centering
  \setlength{\fboxsep}{0pt} 
  \includegraphics[width=0.75\textwidth]{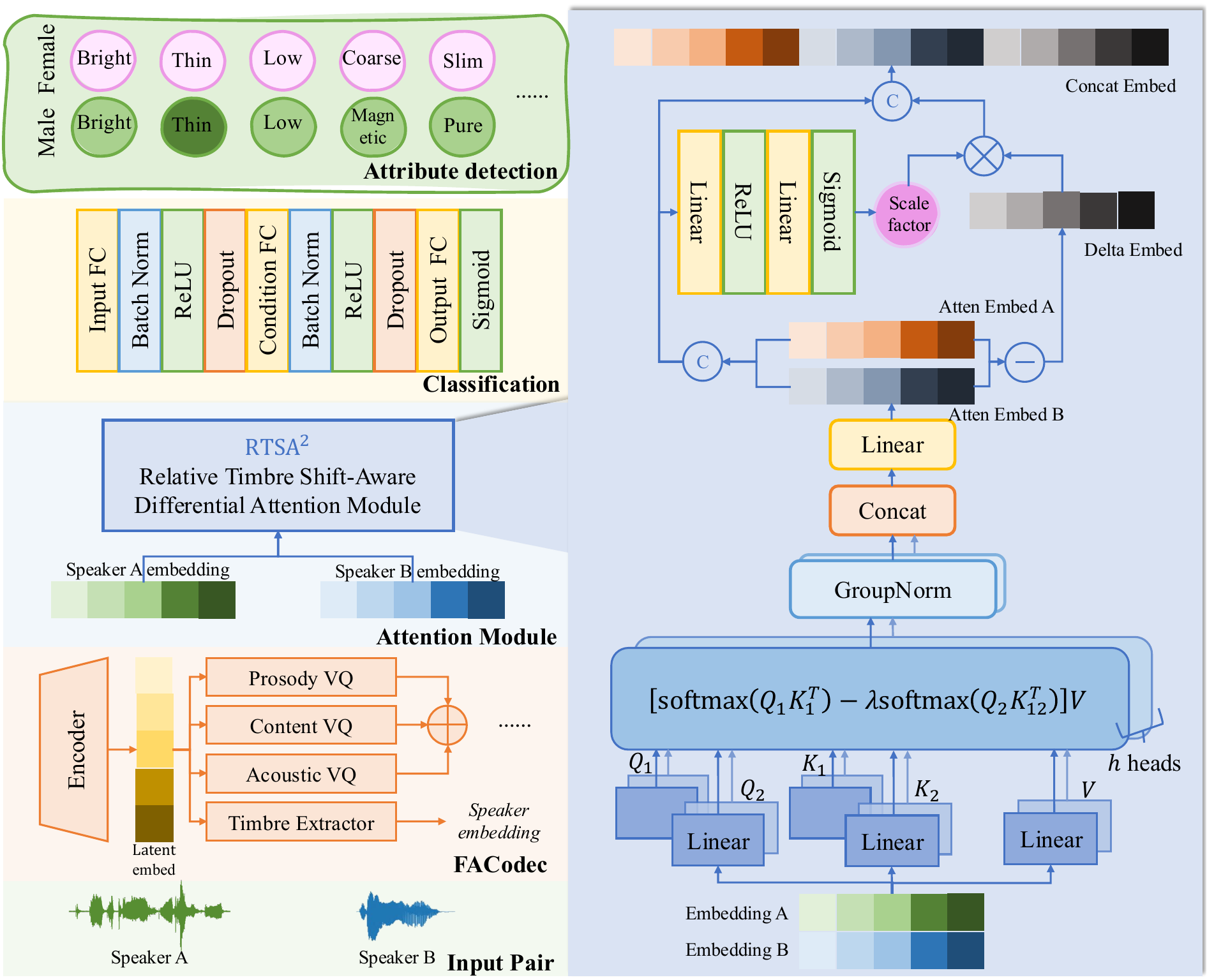}
  \caption{Overview of the proposed QvTAD framework.}
  \label{fig:main-method}
\end{figure}

\section{Method}

This paper focuses on the task of modeling the relative strength of timbre-related attributes in speech, aiming to determine which of two given speech samples exhibits a stronger presence of a target timbre attribute. To this end, we propose QvTAD, a novel model that integrates a differential attention mechanism with a perceptual difference amplification strategy for relative attribute prediction. Our approach follows the paradigm of “sample pair + attribute word + relative strength judgment” and is composed of a pairwise comparison oriented data augmentation method with Disjoint-Set Searching and a main prediction model with two core components: (1) a Differential Attention Prior-Guided Relative Attribute Amplification Mechanism, which enhances the perception of fine-grained attribute differences between sample pairs, and (2) a multi-label classifier conditioned on the target attribute. As illustrated in~\ref{fig:main-method}, the overall architecture comprises three main modules: a frozen FACodec Encoder for timbre representation extraction, a differential attention module for contrastive modeling, and a final probabilistic prediction network with sigmoid activation.

\subsection{Data Augmentation with Disjoint-Set Searching}

We analyzed an existing speech attribute comparison dataset (see Section~\ref{dataset} for VCTK details), which includes multiple attribute dimensions (e.g., brightness, roughness) defined by speaker pairwise comparisons.

Statistical analysis reveals a pronounced long-tail distribution: head attributes (e.g., \textit{Brightness\_F}, \textit{Rough\_F}) account for over 45\% of samples, while tail attributes (e.g., \textit{Transparent\_F}, \textit{Gentle\_F}) are severely underrepresented (some <1\%). This imbalance may bias training, hindering the learningof rare attributes.

To address the issue of imbalanced data distribution across timbre attributes and insufficient coverage of training pairs, we propose a Disjoint-Set Union (DSU)-based data augmentation strategy tailored for pairwise comparison tasks. Inspired by structured modeling, our method abstracts each speaker and their associated timbre attribute strength as nodes and edges within a directed graph, enabling the construction of an attribute-aware graph structure that systematically discovers potential, yet unobserved, comparable pairs to enrich the training data.

Specifically, each annotated pair in the training set is represented as a directed edge between two nodes of the form \textit{⟨Speaker, Attribute⟩}. For example, if utterance $\mathcal{O}_B$ is annotated as stronger than $\mathcal{O}_A$ with respect to attribute $v_i$, we add a directed edge from node $(A, v_i)$ to $(B, v_i)$, indicating the relative strength relation. All annotated pairs form a directed graph under each attribute dimension. Under the assumption of transitivity in attribute comparison, this graph is expected to form a locally ordered DAG.

Building on this structure, we introduce an attribute-specific DSU-based augmentation strategy. For each attribute $v_i \in \mathcal{V}$, we maintain an independent disjoint-set structure to track transitive connectivity among nodes under that attribute. We then iterate over all unannotated node pairs $(u, v)$ and query their connectivity in the DSU: if a transitive path exists between $u$ and $v$ but the pair is not present in the original training set, the pair is treated as a pseudo-labeled candidate pair. To mitigate the risk of noisy or inconsistent annotations inherent to graph construction, we apply a minimum path length threshold and a multi-path voting mechanism to ensure the reliability and quality of the augmented pairs.

\subsection{Relative Timbre Shift-Aware Differential Attention}

\subsubsection{Input Modeling and Task Formulation}

We consider as input a pair of speech embeddings, denoted as $\mathbf{e}_A, \mathbf{e}_B \in \mathbb{R}^d$, where $d = 256$ is the embedding dimensionality. The timbre attribute of interest is specified by an attribute index vector $\mathbf{l} \in \{0, 1\}^K$, where $K = 34$ denotes the number of predefined timbre attributes. The objective is to predict whether $\mathbf{e}_B$ exhibits a stronger presence of the specified attribute compared to $\mathbf{e}_A$. The model outputs a scalar probability $p \in [0, 1]$, representing the confidence that "sample B is stronger than sample A" for the given attribute.

Following NaturalSpeech3~\cite{naturalspeech3}, we employ a pretrained FACodec Encoder to extract 256-dimensional speech embeddings for each audio sample. The encoder is frozen during training to maintain stable representations.

\subsubsection{Differential Attention Prior-Guided Relative Attribute Amplification}

Given the nature of relative attribute modeling—centered around contrastive perception—we introduce a differential attention mechanism to enhance representation differences between samples. This module employs multi-head differential denoising attention~\cite{diff-attention} to suppress shared noise in the attention maps between the query-key pairs, thereby strengthening discriminative contrast features.

We begin by concatenating the two input embeddings into a sequence:

\begin{equation}
\mathbf{E} = [\mathbf{e}_A;\ \mathbf{e}_B] \in \mathbb{R}^{2 \times d}
\label{eq:concat_input}
\end{equation}

This sequence is fed into the multi-head differential attention module to 
generate enhanced contrastive representations: $ \mathbf{e}_A^{\text{att}},\ \mathbf{e}_B^{\text{att}} \in \mathbb{R}^d $. Inspired by differential amplifiers, the module computes attention differences between two sets of query-key pairs to suppress common-mode signals and amplify significant contrastive cues. Specifically, each attention head computes:

\begin{equation}
[Q_1;\ Q_2] = \mathbf{E} W_Q,\quad [K_1;\ K_2] = \mathbf{E} W_K
\label{eq:qk_proj}
\end{equation}

The attention difference is then computed as:

\begin{equation}
\text{DiffAttn}(\mathbf{E}) = \text{softmax}\left(\frac{Q_1 K_1^\top}{\sqrt{d}}\right) - \lambda \cdot \text{softmax}\left(\frac{Q_2 K_2^\top}{\sqrt{d}}\right)
\label{eq:att_diff}
\end{equation}

where $\lambda \in (0, 1)$ is a learnable scaling factor (initialized as $\lambda_{\text{init}}$) to balance the contribution of the two attention maps.

Since relative attribute modeling always operates on a pair of samples, the input sequence length remains fixed at 2. As such, our model focuses on pairwise contrast rather than long-range contextual dependency, and thus positional encoding is unnecessary. Furthermore, we empirically observed that Rotary Positional Embeddings (RoPE) did not improve performance in this setting and introduced unnecessary computation. We therefore exclude RoPE from the attention head to simplify the model and improve training efficiency.

To further enhance the model’s sensitivity to subtle timbre differences, we compute a difference vector, which is non-linearly modulated and normalized to generate an amplified contrastive representation:

\begin{equation}
\Delta = \mathbf{e}_B^{\text{att}} - \mathbf{e}_A^{\text{att}}; 
\hat{\Delta} = \tanh(\Delta) \cdot \|\Delta\|_2 \cdot \gamma
\label{eq:delta}
\end{equation}

where $\gamma \in [0, 2]$ is a learnable amplification factor that scales the perceptual contrast based on the difficulty of the sample pair. We empirically found this bounded range leads to stable training and better performance. The amplification factor $\gamma$ is generated by a contrastive prediction module:

\begin{equation}
\gamma = \sigma\left(f_{\text{scale}}([\mathbf{e}_A^{\text{att}};\ \mathbf{e}_B^{\text{att}}])\right)
\label{eq:gamma}
\end{equation}

where $f_{\text{scale}}(\cdot)$ is a two-layer feedforward network with non-linear activations, and $\sigma(\cdot)$ denotes the sigmoid function. The output is rescaled to the range $[0, 2]$ to control the intensity of contrast amplification.

Finally, we concatenate three sources of information—$\mathbf{e}_A^{\text{att}}$, $\mathbf{e}_B^{\text{att}}$, and $\hat{\Delta}$—and feed them into a feed-forward prediction network composed of two fully connected layers, batch normalization, and dropout for regularization. The output is a probability vector:

\begin{equation}
\mathbf{o} = \sigma(W{\text{out}} \mathbf{z}) \in \mathbb{R}^K
\label{eq:output}
\end{equation}

where the $k$-th element $o_k$ represents the predicted probability that sample B is stronger than sample A with respect to attribute $k$. During training, we apply supervision only on the target attribute dimension where $l_k = 1$, using a binary cross-entropy (BCE) loss.

\section{Experiments and Evaluation}

\subsection{Dataset}
\label{dataset}

For the vTAD task, we utilize the VCTK-RVA dataset, which is derived from the publicly available CSTR VCTK Corpus \cite{vctk} through additional annotation. The original VCTK corpus consists of speech recordings from 110 English speakers with diverse accents. Each speaker reads approximately 400 sentences, selected from a newspaper article, the Rainbow Passage, and an elicitation paragraph used in the Speech Accent Archive.

VCTK-RVA provides relative strength annotations for voice attributes between speech segments from different speakers of the same gender. The training set involves 101 speakers. For each ordered pair of speakers, 1 to 3 timbre descriptors are annotated, resulting in a total of 6,038 labeled pairs in the form of {Speaker A, Speaker B, voice attribute v}, indicating that Speaker B exhibits a stronger presence of attribute v compared to Speaker A.

After data augmentation, the training set contains 166,409 samples. For validation and testing, we use the original splits provided with the dataset. The validation set consists of 1,085 samples. The test set is divided into two subsets: seen (94,000 samples) and unseen (91,600 samples), representing speakers included and excluded from the training set, respectively. All audio samples are downsampled to 16 kHz.

\subsection{Training Setup}
For fair evaluation on unseen speakers, we adopt the pretrained codec model in NaturalSpeech3, FACodec as the timbre encoder which produces 256-dimensional representations. FACodec is trained on the Librilight corpus, which ensures no speaker overlap with the VCTK test set.

Model training is performed on 8 NVIDIA A800 GPUs via the Accelerate framework~\cite{accelerate}, with a batch size of 8,192 samples per GPU. The model is trained for 20 epochs until convergence. We employ the Adam optimizer with hyperparameters $\beta_1 = 0.9$, $\beta_2 = 0.999$, and $\epsilon = 1 \times 10^{-8}$, with a learning rate of $5 \times 10^{-5}$.

For baseline evaluation, we follow the official configuration provided by the vTAD Challenge, using both the dataset and training/inference protocols as described in the challenge paper and official codebase \footnote[1]{https://github.com/vTAD2025-Challenge/vTAD}.

\subsection{Evaluation and Ablation Analysis}
In this section, the QvTAD model is compared with the reproduced results of the official FAcodec baseline model of the vTAD 2025 challenge~\cite{challenge}, with the original data from the paper listed above. Following the evaluation framework proposed in~\cite{vTAD-dataset}~\cite{challenge}, accuracy (ACC) and equal error rate (EER) are adopted as the evaluation metrics, and a systematic comparison is conducted on the performance of different models across two gender categories (Male, Female) and two scenarios (Seen, Unseen). Among them, QvTAD-AST corresponds to the model data of the competition scheme, which is included solely to ensure the fairness of the comparison; QvTAD-RTSA$^2$ represents the proposed model in this paper.
The experimental results demonstrate that in the Seen scenario, QvTAD-RTSA$^2$ achieves the second-highest accuracy for both genders, with its performance surpassed only by QvTAD-AST and FACodec (Reproduced). In the Unseen scenario, the proposed model attains the best performance across all evaluation metrics, fully validating the superior capability of the proposed method in learning speaker-irrelevant differences.

\begin{table*}[t]
\caption{Performance comparison across different speaker groups on VCTK-RVA testset.}
\label{tab:seen_unseen}
\centering
\setlength{\tabcolsep}{3pt} 
\renewcommand{\arraystretch}{1.2} 

\begin{tabular}{l|ccr|ccr}
\toprule
\multirow{2}{*}{\textbf{Method}} & 
\multicolumn{3}{c|}{\textbf{Seen (M/F)}} & 
\multicolumn{3}{c}{\textbf{Unseen (M/F)}} \\
\cmidrule(lr){2-4} \cmidrule(lr){5-7}
& ACC (\%) $\uparrow$ & EER (\%) $\downarrow$ & Avg ACC &
  ACC (\%) $\uparrow$ & EER (\%) $\downarrow$ & Avg ACC \\
\midrule
ECAPA-TDNN (Reported) & 
94.23/93.42 & 5.43/6.45 & 93.83 &
71.01/70.18 & 27.10/30.28 & 70.60 \\

FACodec (Reported) & 
92.67/93.37 & 6.85/6.58 & 93.02 &
91.67/89.77 & 8.72/10.38 & 90.72 \\

\midrule

\arrayrulecolor{black!30}

FACodec (Reproduced) & 
\textbf{87.49}/84.57 & \textbf{12.17}/\uline{14.92} & \uline{86.03} &
\uline{81.55}/70.42 & 18.04/29.66 & \uline{75.99} \\

\textsc{QvTAD-AST} & 
84.95/\textbf{88.79} & 15.55/\textbf{12.03} & \textbf{86.87} &
74.00/\uline{76.44} & 26.59/\uline{25.30} & 75.22 \\

\textsc{QvTAD-RTSA$^2$} & 
\uline{86.45}/\uline{85.32} & \uline{13.82}/14.93 & 85.89 &
\textbf{91.90}/\textbf{82.07} & \textbf{7.89}/\textbf{18.52} & \textbf{86.99} \\

\bottomrule
\end{tabular}
\end{table*}

To evaluate the effectiveness of each proposed component in our model, we conduct ablation experiments by progressively removing key modules and comparing performance on both seen and unseen speaker groups. Table~\ref{tab:ablation} summarizes the results.

We first evaluate the impact of our DSU Searching-based data augmentation. Removing it leads to significant performance degradation, particularly for seen speakers (-2.12\% ACC) and to a lesser extent for unseen speakers (-1.44\% ACC). This confirms that augmentation enhances the model's robustness to intra- and inter-speaker variations, promoting better generalization.
Next, we assess our RTSA$^2$ module, designed to accentuate perceptual differences between embedding pairs while suppressing shared timbral features. Removing this module causes a slight drop in accuracy for unseen speakers (-0.71\%) despite a marginal gain for seen speakers (+0.10\%), suggesting RTSA$^2$ is critical for generalizing to unseen speaker distributions. This is likely attributed to its contrastive learning mechanism, which emphasizes relative timbre shifts over absolute representations.

Overall, the full QvTAD-RTSA$^2$ model achieves the best performance across both seen and unseen conditions, validating the effectiveness of both components. The results highlight the importance of contrast-aware attention and data diversity in relative timbre attribute modeling.

\begin{table}[!htbp]
    \centering
    \vspace{-2\baselineskip}
    \setlength{\tabcolsep}{5pt}
    \renewcommand{\arraystretch}{1.2}
    \caption{
    Ablation study of module effectiveness between seen and unseen speaker groups. 
    }
    \label{tab:ablation}
    \begin{tabular}{l|cc}
        \toprule
        \textbf{Method} & \textbf{Seen ACC (\%)} & \textbf{Unseen ACC (\%)} \\
        \midrule
        QvTAD-RTSA$^2$ & 85.89 & 86.99 \\
        w/o Data Augmentation & 83.77 & 85.55 \\
        w/o RTSA$^2$ Module & 85.99 & 86.28 \\
        \bottomrule
    \end{tabular}
\end{table}

\vspace{-24pt}

\section{Conclusion}
We introduce QvTAD, a novel differential attention-based framework for modeling voice timbre attributes through pairwise comparisons. By integrating a graph-based data augmentation method and a shift-aware differential attention mechanism, QvTAD effectively addresses the challenges of label imbalance and perceptual subjectivity in timbre modeling. Experiments on the VCTK-RVA dataset demonstrate that QvTAD achieves superior performance across a range of timbre descriptors, especially in cross-speaker evaluation. These results highlight the potential of comparative comparison of voice timbre attributes and fine-grained acoustic modeling for advancing perceptual attribute understanding in speech generation. Future work may extend this framework to multi-lingual timbre analysis or real-time controllable synthesis.


%

\bibliographystyle{splncs04}
\bibliography{vtad}

\end{document}